\documentclass[modern]{aastex61}
\usepackage{graphicx}
\usepackage{amsmath}
\usepackage{amssymb}
\usepackage{enumitem}

\newcommand\mybar{\kern1pt\rule[-\dp\strutbox]{.8pt}{\baselineskip}\kern1pt}

\setlist[itemize]{noitemsep, topsep=0pt, leftmargin=*}

\shorttitle{Graviton Mass}
\shortauthors{Loeb}

\begin{document}

\title{A New Limit on the Graviton Mass from the Convergence Scale of
  the CMB Dipole}

\author{Abraham Loeb}
\affiliation{Astronomy Department, Harvard University, 60 Garden
  St., Cambridge, MA 02138, USA}

\begin{abstract}

The clustering dipole in the 2MASS galaxy survey converges on a scale
of $\sim 400~{\rm Mpc}$ to the local peculiar velocity inferred from the
Cosmic-Microwave-Background dipole. I show that this limits the
graviton mass in Yukawa theories of gravity to less than $5\times
10^{-32}~{\rm eV}$. The new limit is $2.5\times 10^8$ times tighter
than the latest constraint from gravitational waves detected by the
LIGO-Virgo-KAGRA collaboration.

\end{abstract}

\bigskip\bigskip

\section{Introduction}

In General Relativity, gravitational interactions are mediated by a
massless gauge boson, the graviton. However, a class of modified
gravity theories endow the graviton with a mass, $m_g$ (for a
comprehensive review, see \citet{2017RvMP...89b5004D}).

The possible existence of a graviton mass causes a time delay in the
propagation of gravitational waves relative to the speed of light, and
could also lead to a Yukawa-form for the gravitational potential
of a point mass $M$ as a function of distance $r$,
\begin{equation}
\Phi = - {GM\over r} \exp(- r/\lambda_g),
\end{equation}
where $G$ is Newton's constant.  This is the Green's function solution
to the screened Poisson equation,
\begin{equation}
(\nabla^2 - \lambda_g^{-2})\Phi= 4\pi GM \delta^3({\bf r}),
\end{equation}
with the gravitational acceleration being ${\bf g}=-{\bf \nabla}\Phi$.

The screening length is given by the Compton wavelength of the
graviton,
\begin{equation}
  \lambda_g={h\over m_g c}= \left({400~{\rm Mpc} \over m_g/10^{-31}~{\rm
      eV}}\right),
  \label{lambda}
\end{equation}
where $h$ is Planck's constant and $c$ is the speed of light.

Many limits of the graviton mass were derived in the literature based
on a variety of data sets (for a comprehensive summary of the limits
and related references, see the ~\citet{PDG2022}).  Most recently,
analysis of cosmological gravitational-wave data from the three
observing runs of the LIGO-Virgo-KAGRA collaboration (GWTC-3), used
the propagation speed limit to derive the constraint, $m_g < 1.27
\times 10^{-23}~{\rm eV}$~\citep{2021arXiv211206861T}. Here, we derive
a tighter cosmological constraint.

\section{Cosmic Dipole Convergence}

Peculiar velocities relative to the Hubble flow, ${\bf v}$, are
sourced by cosmic density inhomogeneities, with ${\bf v}\propto {\bf
  g}$ in the linear regime~\citep{2011ApJ...741...31B}. The possible
screening of gravity on a spatial scale $\lambda_g$ introduces an
exponential cutoff to the gravitational influence of mass
concentrations beyond that scale. In that case, the net gravitational
acceleration can be written as,
\begin{equation}
  {\bf g}({\bf r})=G\int d{\bf r'} {\rho_m({\bf r'})\exp(-\vert {\bf
      r'}-{\bf r}\vert/\lambda_g) \over \vert {\bf r'}-{\bf
      r}\vert}\left({1 \over \vert {\bf r'}-{\bf r}\vert}+{1\over
    \lambda_g}\right) {({\bf r'}-{\bf r})\over \vert {\bf r'}-{\bf
      r}\vert},
\label{horizon}
\end{equation}
where $\rho_m ({\bf r'})$ is the cosmic matter density at comoving
position ${\bf r'}$.

The peculiar velocity of the Local Group of galaxies was measured
through the dipole anisotropy of the Cosmic Microwave Background
(CMB)~\citep{2014A&A...571A..27P}. The matter perturbations mapped by
the 2MASS galaxy survey were shown to converge to the CMB dipole (to
within one standard deviation, corresponding to $\sim 10\%$ of the
measured CMB dipole) at a distance of $\sim 400$ Mpc (see Figure 7
in~\citet{2011ApJ...741...31B}).

\section{Graviton Mass Limit}

The requirement that the exponential suppression of a massive graviton
would not spoil the 2MASS dipole convergence by more than one standard
deviation, sets the constraint $\lambda_g \gtrsim 800$ Mpc based on
equation~(\ref{horizon}). Equation~(\ref{lambda}) therefore implies,
\begin{equation}
m_g< 5\times 10^{-32}~{\rm eV},
\end{equation}
or equivalently $m_g<8.9\times 10^{-65}$~g.

This limit is tighter by a factor of $2.5\times 10^8$ than the
LIGO-Virgo-KARGA limit~\citep{2021arXiv211206861T}, and constitutes
the best Yukawa-limit on the graviton mass so far~\citep{PDG2022}.

\bigskip
\bigskip
\bigskip
\bigskip
\section*{Acknowledgements}

This work was supported in part by Harvard's {\it Black Hole
  Initiative}, which is funded by grants from JFT and GBMF.

\bigskip
\bigskip
\bigskip

\bibliographystyle{aasjournal}
\bibliography{t}
\label{lastpage}
\end{document}